\documentclass[
    reprint,         
    aps,             
    prl,             
    superscriptaddress, 
    showpacs,        
    floatfix,        
    nofootinbib      
]{revtex4-2}

\usepackage[utf8]{inputenc}    
\usepackage[T1]{fontenc}       
\usepackage{amsmath, amssymb}  
\usepackage{bm}                
\usepackage{graphicx}          
\usepackage{xcolor}            
\usepackage{hyperref}          
\usepackage{physics}           
\usepackage{siunitx}           
\usepackage{microtype}         
\usepackage{newtxtext}         
\usepackage{newtxmath}         
\usepackage{multirow}          
\usepackage{threeparttable}    
\usepackage[version=3]{mhchem} 
\usepackage{booktabs}
\hypersetup{
    colorlinks=true,
    linkcolor=blue,
    citecolor=blue,
    urlcolor=blue,
    pdfauthor={Your Name},
    pdftitle={Your Title},
}

\begin{document}

\title{Direct three-body dynamics govern ion–atom recombination and barrierless termolecular reactions}

\author{Rian Koots}
\affiliation{Department of Physics and Astronomy, Stony Brook University, Stony Brook, New York, New York 11794, USA}

\author{Marjan Mirahmadi}
\affiliation{Max-Born-Institut, Max-Born-Str. 2A, 12489 Berlin, Germany}

\author{Jes\'us P\'erez-R\'ios}
\email{jesus.perezrios@stonybrook.edu}
\affiliation{Department of Physics and Astronomy, Stony Brook University, Stony Brook, New York, New York 11794, USA}

\date{\today}

\begin{abstract}
    For over a century, termolecular (third-order) chemical reactions have been explained by the Lindemann-Hinshelwood mechanism, assuming sequential stabilization via bimolecular encounters. Here, we demonstrate that barrierless termolecular reactions are fundamentally governed by direct three-body dynamics. Using classical trajectory calculations in hyperspherical coordinates, we quantitatively reproduce ion–atom recombination kinetics across a wide temperature range without invoking intermediate complexes or steady-state assumptions. Our results not only resolve longstanding discrepancies between theory and experiment, but also establish a general mechanistic framework for barrierless termolecular reactions, with implications spanning atmospheric chemistry, plasma physics, and ultracold chemistry. 
\end{abstract}

\maketitle

\section{Introduction}

Termolecular or third-order chemical reactions \ce{A + B + C -> Products} are ubiquitous in nature\footnote{It is important to emphasize that in this work, we treat termolecular reactions as third-order reactions, independently of the reaction mechanism behind them.}, from plasma physics to the fate of a single ion in an ultracold atomic bath, thus spanning over 10 decades of temperature and covering a multitude of essential systems in physics and chemistry. Termolecular reactions are often the only possible route to a product, such as for ozone formation in the stratosphere through \ce{O + O2 + M -> O3 + M}, where M is generally \ce{N2}, or the formation of primordial stars where molecular hydrogen, required as a coolant, appears as a consequence of the reaction \ce{H + H + H -> H2 + H}. 

Trautz and Bodenstein pioneered the study of termolecular reactions~\cite{Trautz_1914,Trautz_1916,Trautz_1916_bis,Bodenstein_1922}, both of whom were the main figures in the development of modern chemical kinetics. As early as 1914, Max Trautz, after studying NO + NO + Cl $\rightarrow$ NOCl + NOCl, and developing the first ideas on molecular collision theory, proposed that this third-order reaction occurred sequentially, as two concatenated bimolecular reactions~\cite{Trautz_1914}. The reasoning was that it is easier for two molecules to find each other than for three molecules to find each other simultaneously. In 1922, Bodenstein, studying a different third-order chemical reaction, proposed the same mechanism to explain its kinetics~\cite{Bodenstein_1922}. Bodenstein's reasoning was very similar to that of Trautz. However, these mechanisms were rather speculative and lacking any theoretical foundation, except for Bodenstein, who proposed a model for the third-order reaction rate. The idea of a sequential mechanism was finally formalized by Lindemann and Hinshelwood~\cite{LINDEMANN196793,Hinshelwood1926OnTT}, yielding the so-called Lindemann-Hinshelwood mechanism for termolecular and unimolecular reactions. 

The Lindemann-Hinshelwood mechanism assumes that a third-order reaction, such as 
\begin{align}
\label{r1}
\ce{ A + B  + M \rightarrow AB + M },
\end{align}
occurs sequentially. First, two of the reactants meet to form a complex as 
\begin{align}\label{eq:2step}
\ce{ A + B &<-->[k_2][k_{diss}] AB^* },
\end{align}
where $k_2$ denotes the rate of formation of AB$^*$ complexes and $k_{\text{diss}}$ stands for its dissociation rate. In a second step, the third reactant stabilizes the complex
\begin{align}
\ce{ AB^* + M &->[k_{stab}] AB + M },
\end{align}
where $k_{\text{stab}}$ refers to the stabilization rate due to a collision with a third reactant. Assuming that complex formation reaches a steady state, the termolecular reaction rate of the reaction \ref{r1} is given by
\begin{equation}\label{eq7}
    k_3=\frac{k_2k_{\text{stab}}}{k_\text{diss}+k_{\text{stab}}[\text{M}]},
\end{equation}
where [M] is the number density of the particle M. Therefore, the nature of the intermediate complex and the concentration of the third reactant establish the outcome of the reaction. This argument is compelling to the point that the community has been using this and variants to deal with termolecular reactions ever since~\cite{Xie_Poirier_Gellene_2003,Russell_1985,Russell_1986,Hauser2015,Troe2005,Herbst1980,Bates1979,Brahms2011,Bates1979bis,Luther2005,RBCmechanism}. However, despite its widespread use, the Lindemann–Hinshelwood mechanism fails to quantitatively describe many barrierless recombination processes, particularly at low temperatures~\cite{Perez-Rios2018,Koots2025,Sulfur}. This raises a fundamental question: are barrierless termolecular reactions inherently sequential, or are they governed by direct many-body dynamics?

On the other hand, it is possible that a termolecular reaction indeed occurs in a single step, with the three reactants meeting almost simultaneously (three-body collision), known as the direct mechanism. In principle, in light of the arguments of Trautz, Bodenstein, Lindemann, and Hinshelwood, such a reaction mechanism should be discarded, since a three-body collision is highly unlikely. Nevertheless, the argument does not account for the fact that most termolecular reactions are barrierless and often exhibit inverse-temperature scaling characteristic of long-range capture dynamics. In contrast, most bimolecular reactions require the reactants to overcome an energy barrier to form products. Thus, termolecular reactions are more efficient at low temperatures, whereas bimolecular reactions are at high temperatures. As a consequence, depending on the reaction and the system's physical conditions, the direct approach is unavoidable for barrierless termolecular reactions. Indeed, the Lindemann-Hinshelwood mechanism has already been shown to fail to describe halogen and rare gas atom recombination~\cite{Koots2025}, or ion-atom recombination reactions between cold ions and ultracold atoms~\cite{Perez-Rios2018,Mirahmadi_2022}. 

Ion-atom recombination \ce{A+ + A + A \rightarrow Products} is one of the most studied termolecular reactions, especially \ce{He+ + He + He \rightarrow He2+ + He} and \ce{Ar+ + Ar + Ar \rightarrow Ar2+ + Ar}, both experimentally and theoretically. This reaction is essential in cold chemistry since it establishes the survival of a single ion in a bath of ultracold atoms, or in excimer laser physics~\cite{Excimer}. In plasma physics, ion-atom recombination can affect the yield of molecular ions and hence affect the momentum transfer of the plasma to the walls of the reactor. The vast majority of theoretical efforts to understand these reactions have been influenced by the sequential approach underlying the Lindemann-Hinshelwood mechanism. The treatments, even at the full quantum level, fail to explain the experimental findings. As we demonstrate in this paper, it is indeed the direct mechanism that correctly explains ion-atom recombination kinetics. Our treatment uses classical trajectory calculations in hyperspherical coordinates without invoking steady-state assumptions or intermediate complexes. Our results quantitatively explain the experimental data for ion-atom recombination reactions over a wide range of temperatures, describing them in terms of the underlying potential energy surface. Moreover, our results support previous attempts to extrapolate experimental data into the cold regime, which are otherwise questionable based on theoretical treatments that employ the Lindemann-Hinshelwood mechanism. These findings redefine the mechanistic description of barrierless chemical processes in diverse environments, including ozone formation in the atmosphere, sulfur recombination in geochemical cycles, ion chemistry in plasmas, and molecular ion formation in cold environments.

\section{Methodology}
Here, we briefly review the method used for obtaining the temperature-dependent ion-atom recombination rate coefficients, which has been thoroughly described in Refs.~\cite{Perez-Rios_2014,Mirahmadi_2022}. We use a classical trajectory method that relies on a coordinate transformation from the Cartesian 3D space to a hyperspherical 6D (after removing the center of mass degrees of freedom) space in order to define the three-body recombination cross section as
\begin{equation}\label{eq: sigma}
    \sigma_{\text{rec}}(E_c) = \frac{8\pi^2}{3}\int_0^{b_{\text{max}}(E_c)}\mathcal{P}_{\text{rec}}(E_c,b)b^4db ,
\end{equation}
where $E_c$ represents the collision energy, $\mathcal{P_{\text{rec}}}$ is the opacity function which defines the probability of recombination for the initial conditions $(E_c, b)$, and $b$ is the impact parameter which is now defined as a 5-dimensional vector embedded in the 6D space. 

\begin{figure}[h!]
    \centering
    \includegraphics[width=1\linewidth]{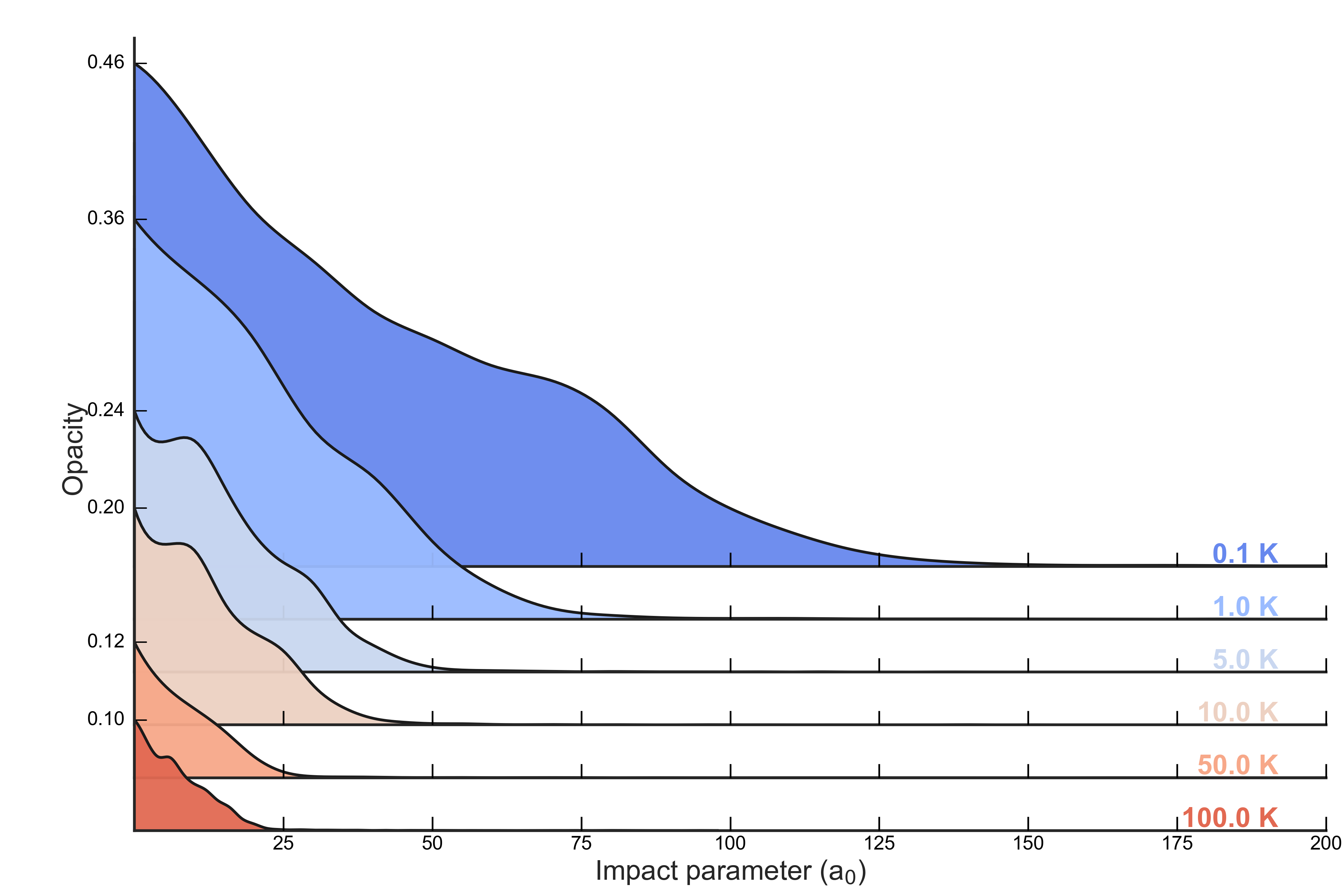}
    \caption{Opacity function of \ce{He+ + He + He -> He2+ + He} for various collision energies. Lower collision energies result in larger values of $b_{\mathrm{max}}$. The cross section is evaluated via the integral of these curves, as shown in Eq.~\ref{eq: sigma}.} 
    \label{fig:opacity_ridgeplot}
\end{figure}

The opacity function for several collision energies is plotted in Figure~\ref{fig:opacity_ridgeplot}, which visually demonstrates the relationship between the collision energy, the probability of reaction, and the impact parameter. As the collision energy decreases, the probability of the reaction increases, as expected for a barrierless reaction. More importantly, we observe that the maximum impact parameter follows ($b_{\mathrm{max}}(E_c) \propto E_c^{-1/4}$) below $E_c = 10$~K, accordingly to the classical threshold law~\cite{Perez-Rios2015} as a result of the preponderance of the charge-induced dipole interaction (ion-atom) over the induced dipole-induce dipole interaction (atom-atom). This integral is evaluated via the Monte Carlo technique, which involves random sampling of the trajectory space over an appropriate range of impact parameters up to the maximum impact parameter, $b_{\text{max}}$.

The thermally averaged recombination rate coefficient is easily obtained by integrating the energy-dependent cross section over the three-body Maxwell-Boltzmann energy distribution as
\begin{equation}
    k_{\text{rec}}(T) = \frac{1}{2(k_bT)^3}\int_0^\infty \sqrt{\frac{2E_c}{\mu}}\sigma_{\text{rec}}(E_c)E_c^2e^{-E_c/(k_bT)}dE_c
\end{equation}
where $k_b$ is the Boltzmann constant, $T$ is the temperature of the system, and $\mu = \sqrt{m_1m_2m_3/(m_1+m_2+m_3)}$ is the three-body reduced mass.

For \ce{He2+} recombination, we include a factor of $1/2$ due to the symmetry-induced isotope effect arising from collisions of the same isotope~\cite{gelleneSymmetrydependentIsotopic1993}. For \ce{Ar2+} recombination, we use a factor of $2/3$ to account for the splitting of \ce{Ar2+} electronic states. Potential energy surfaces calculated with spin-orbit coupling, where four out of the six states are attractive~\cite{GADEA1996281}.

All the calculations have been performed using Py3BR software~\cite{Py3BR} running batches of 50,000 trajectories per impact parameter and including up to 60 impact parameters per collision energy, yielding about $10^6$ trajectories per collision energy. The initial hyperradius $R$ for \ce{He2+} recombination was chosen as $500 \pm 50$ a$_0$, and $5,000 \pm 500$ a$_0$ for \ce{Ar2+} recombination.

\subsection{Potential energy surfaces}\label{PES}
It has been shown that for atom and ion-atom recombination reactions the role of the three-body interaction term is negligible~\cite{Yu_2024}. Thus, we use a pairwise additive potential to describe the interaction among the three colliding atoms, such that $V(r_1,r_2,r_3) = V(r_{12}, r_{23}, r_{31})$. The \ce{He2+} interaction is described by the following potential of Ref.~\cite{Chang_Gellene_2003}, whereas \ce{He2} is described by the analytic Hartree-Fock-dispersion potential HFD-B3-FCI1~\cite{Aziz_Slaman_1990}. In addition we use Lennard-Jones potentials ($LJ(r) = C_m/r^m - C_n/r^n$) to explore the role of the short-range interactions. For the ion-induced dipole interaction $LJ_{He_{2}^+}(r)$, we use $C_4 = 0.69$ a.u., $C_8 = 1.32$ a.u., and for the van-der Waals interaction $LJ_{{He_2}}(r)$, we use $C_6 = 1.35$ a.u., $C_{12} = 1.38\times 10^4$ a.u.. 

For ion-atom recombination in argon, we employed Lennard-Jones potentials for the atom-atom interaction ($C_6 = 63.5$ a.u., $C_{12} = 2.07\times 10^6$ a.u.) and the ion-atom interaction ($C_{4} = 43.87$ a.u., $C_{8} = 9434$ a.u.), where the short-range interaction follows $r^{-8}$. More details about the six model potentials are given in Table~\ref{tbl:potentials}.

\begin{table}[h]
  \centering
  \small
  \caption{Potential parameters used for ion-atom recombination in helium and argon, with  the dissociation energy ($D_e$) and equilibrium distance ($r_e$) across models. All parameters are in atomic units.}
  \label{tbl:potentials}
  
  \begin{tabular}{@{}l l cc@{}}
    \toprule
    \textbf{Type} & \textbf{Model} & $D_e$ & $r_e$ \\
    \midrule
    Ion - Neutral & $V_{\ce{He2+}}$  & 0.09027 & 2.0473 \\
                  & $LJ_{\ce{He2+}}$ & 0.09017 & 1.3993 \\
                  & $LJ_{\ce{Ar2+}}$ & 0.05099 & 4.5532 \\
    \midrule
    Neutral - Neutral  & $V_{\ce{He2}}$   & $3.4695 \times 10^{-5}$ & 5.6094 \\
                        & $LJ_{\ce{He2}}$  & $3.3016 \times 10^{-5}$ & 5.2287 \\
                        & $LJ_{\ce{Ar2}}$  & $4.8722 \times 10^{-4}$ & 6.3435 \\
    \bottomrule
    \end{tabular}
\end{table}


\section{Results and Discussion}
\subsection{Helium ion recombination}
\begin{figure}[h]
    \centering
    \includegraphics[width=\linewidth]{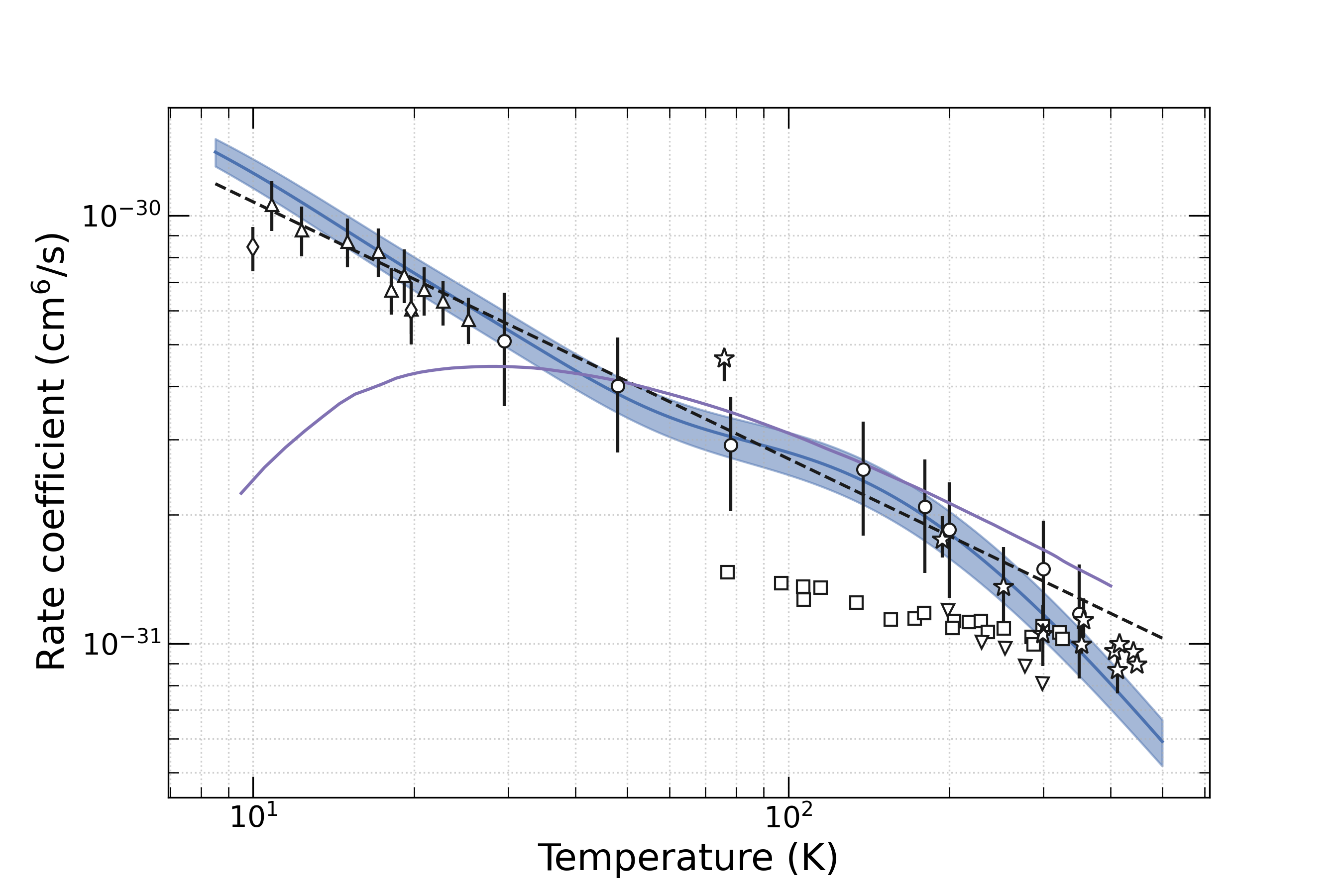}
    \caption{Temperature-dependent formation rate of \ce{He2+} via three-body recombination. The results of the present work, based on the direct mechanism, are shown in blue, with the uncertainty indicated by the shaded region. The blue dashed line represents the classical power-law behavior, $\propto \mathrm{T}^{-3/4}$. The experimentally determined rate coefficients are taken from Ref.~\cite{bohringerTemperatureDependence1983} (circles), Ref.~\cite{johnsenThreebodyAssociation1980a} (squares), Ref.~\cite{nilesTemperatureDependence1965} (x), Ref.~\cite{plasilStabilizationH+H22012} (triangles), Ref.~\cite{jonesTemperatureDependence1980} (inverted triangles), Ref.~\cite{Beaty_Patterson_1965} (stars), and Ref.~\cite{Gerlich_Horning_1992} (diamonds). The black dashed line represents the current standard fit to the experimental data from Refs.~\cite{bohringerTemperatureDependence1983, plasilStabilizationH+H22012, Gerlich_Horning_1992}, given by $1.4\times10^{-31}(300/\mathrm{T})^{0.6}$. The purple curve shows the theoretically predicted rate reported in Ref.~\cite{Xie_Poirier_Gellene_2003}} 
    \label{fig:k_thermal}
\end{figure}
Ion-atom recombination is a barrierless exothermic process, suggesting an increasing trend of the reaction rate coefficient with decreasing temperatures. In Figure~\ref{fig:k_thermal}, we see that this trend is present in the experimentally measured termolecular formation rate of \ce{He2+} between 10~K and 500~K. The temperature-dependent direct recombination rate coefficient is also presented in this temperature range, which accurately captures the experimentally measured data in both temperature trend and magnitude. In addition, we plot the most recent theoretical results from a quantum dynamical study under the Lindemann-Hinshelwood framework~\cite{Xie_Poirier_Gellene_2003}. Here, we see a comparison between the termolecular one-step, direct recombination mechanism and the two-step bimolecular Lindemann mechanism. Although there is general agreement between the two theories above $\sim$30~K, they demonstrate a different behavior at lower temperatures. In particular, the two-step approach presents a maximum near 30~K before rapidly decreasing, which is indicative of an activation barrier in the reaction since still this temperature is far beyond the Wigner threshold limit (constant reaction rate at very low collision energies). On the other hand, the direct mechanism yields a power law behavior where ion atom recombination dominates at low temperatures, as described by the classical threshold law~\cite{Perez-Rios2015} and expected for a barrierless reaction.

The Lindemann-Hinshelwood mechanism fails because not all intermediate states can be considered, due to the complexity of the energy landscape. In practice, the number of complexes and the nature of those considered make a difference. The best example is the results of Smirnov and Dickenson et al., both using the Lindemann-Hinshelwood mechanism but yielding different results for the same ion-atom recombination reaction~\cite{Perez-Rios2018,Perez-Rios2024}. On the contrary, the direct mechanism naturally includes all the most relevant states, thereby improving its description of the chemical kinetics of barrierless termolecular reactions. 

In addition to the quantal results, the current standard for the rate coefficient of \ce{He2+} recombination is based on the best fit to the experimental data of Refs.~\cite{bohringerTemperatureDependence1983,plasilStabilizationH+H22012, Gerlich_Horning_1992}:
\begin{equation}
\label{eq_exp}
    k_{\mathrm{fit}}(T) = 1.4\times 10^{-31}(300/T)^{0.6},
\end{equation}
as shown in Figure~\ref{fig:k_thermal}. The obvious disagreement has raised questions about whether the experimental fitting can be extrapolated to lower temperatures, or if the rate rapidly decreases as the two-step approach suggests~\cite{gerlichInfraredSpectroscopy2018}. Now, we have a third look at this reaction, which agrees with the low-temperature behavior of the experimental data while showing excellent agreement throughout the experimental temperature range. Therefore, the direct mechanism yields reaction kinetics compatible with experimental findings.

\begin{figure}[h]
    \centering
    \includegraphics[width=1\linewidth]{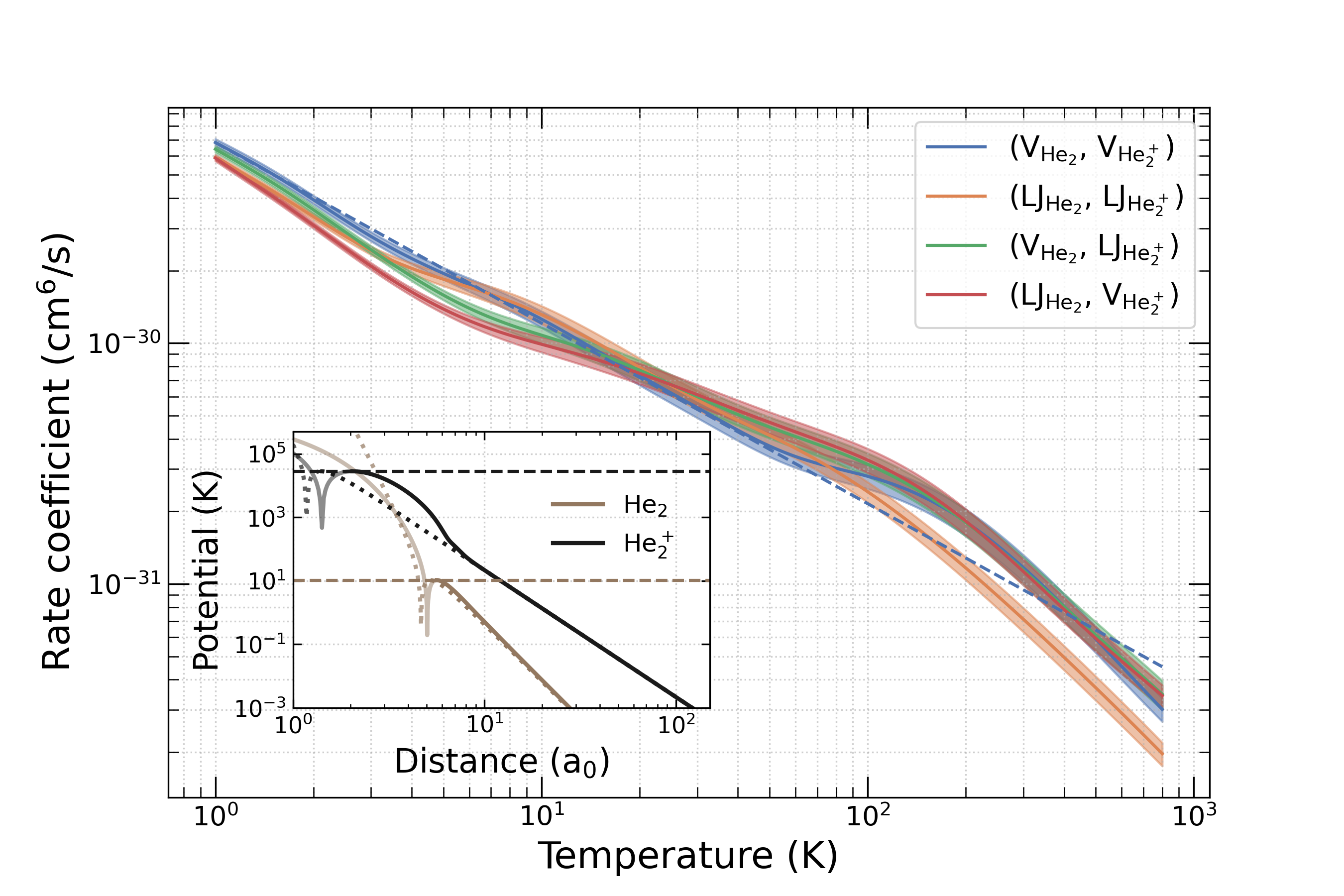}
    \caption{Thermal rate coefficient of \ce{He2+} recombination within the direct approach, calculated with varying potential functions to describe the \ce{He2} and \ce{He2+} interaction. The blue curve is the same as presented in Figure~\ref{fig:k_thermal}, such that the (\ce{He2}, \ce{He2+}) potentials are $(V_{\ce{He2}}$,$V_{\ce{He2+}})$. The orange $(LJ_{\ce{He2}}$,$LJ_{\ce{He2+}})$, green $(V_{\ce{He2}}$,$LJ_{\ce{He2+}})$, and red $(LJ_{\ce{He2}}$,$V_{\ce{He2+}})$ curves are the results of different combinations of potentials. The blue dashed line corresponds to the $T^{-3/4}$ classical threshold law shifted to match the blue curve at low temperature. The inset shows the log-scaled absolute value of the four potentials, $V_{\ce{He2}}$ (solid brown), $V_{\ce{He2+}}$ (solid black), $LJ_{\ce{He2}}$ (dotted brown), $LJ_{\ce{He2+}}$ (dotted black). The dissociation energies of each molecule are shown as horizontal dashed lines in their respective colors, and the repulsive parts of the potentials are 50\% opaque.}
    \label{fig:k3_compare}
\end{figure}

At low temperatures, the direct three-body recombination rate is controlled by the long-range interaction for which the classical threshold law behavior of ion-atom  recombination has been derived within a capture model~\cite{Perez-Rios2015,Mirahmadi_Pérez-Ríos_2023} as $k_3(T) \propto T^{-3/4}$. This behavior occurs in the low-energy regime, or energies below the smallest dissociation energy of the system (in this case $\sim 10$~K). Above this threshold, short-range interactions begin to play a role in the reaction dynamics, leading to deviations from this power-law behavior. This phenomenon explains the discrepancy between the experimentally derived power law [see Eq.~(\ref{eq_exp})] and the classical threshold law for \ce{He2+} formation.

In Figure~\ref{fig:k3_compare}, we investigate the role of the interaction potential on the ion-atom recombination rate coefficient between 1~K and 800~K. The calculations were replicated with varying sets of pairwise potentials, where either one or both of the pairwise interactions are described by the generalized Lennard-Jones potentials described in Section~\ref{PES}. In the inset, the log-scaled absolute value of the four potentials are plotted, which allows for a direct comparison. Below $T = 10$~K, the whole set of potential pairs shows converged rates that follow the classical threshold law behavior. This is expected, since the dynamics are dominated by the $\propto r^{-4}$ charge-induced dipole long-range interaction below collision energies of 10~K, which stays the same independently of the kind of the pairwise potential employed. Therefore, the temperature dependence may be extrapolated to lower temperatures following the power law behavior $k_3(T) \propto T^{-3/4}$. 

As the collision energy increases, the repulsive short-range part of the \ce{He2} interaction comes into play, and the thermal rate deviates from the threshold law. At higher temperatures, slight variations in the pairwise interaction potential (see inset Fig.~\ref{fig:k3_compare}) translate into a different ion-atom recombination rate coefficients. Therefore, the accuracy of our method is such that it is possible to find the best potential that accurately describe the experimental recombination rate. Thus, it is possible to invert the scattering information and transform into an effective potential similar to what it has been done for bimolecular processes~\cite{Aquilanti1988,Aquilanti1999,Aquilanti2002}.

\subsection{Argon ion recombination}
As the lightest rare gas, the formation of \ce{He2+} serves as a benchmark for the study of rare gas ion atom recombination in their parent gas. Here, we extend our calculations to the heavier \ce{Ar2+} recombination reaction to demonstrate that the one-step direct mechanism is responsible for all ion-atom recombination reactions. 

\begin{figure}
    \centering
    \includegraphics[width=1\linewidth]{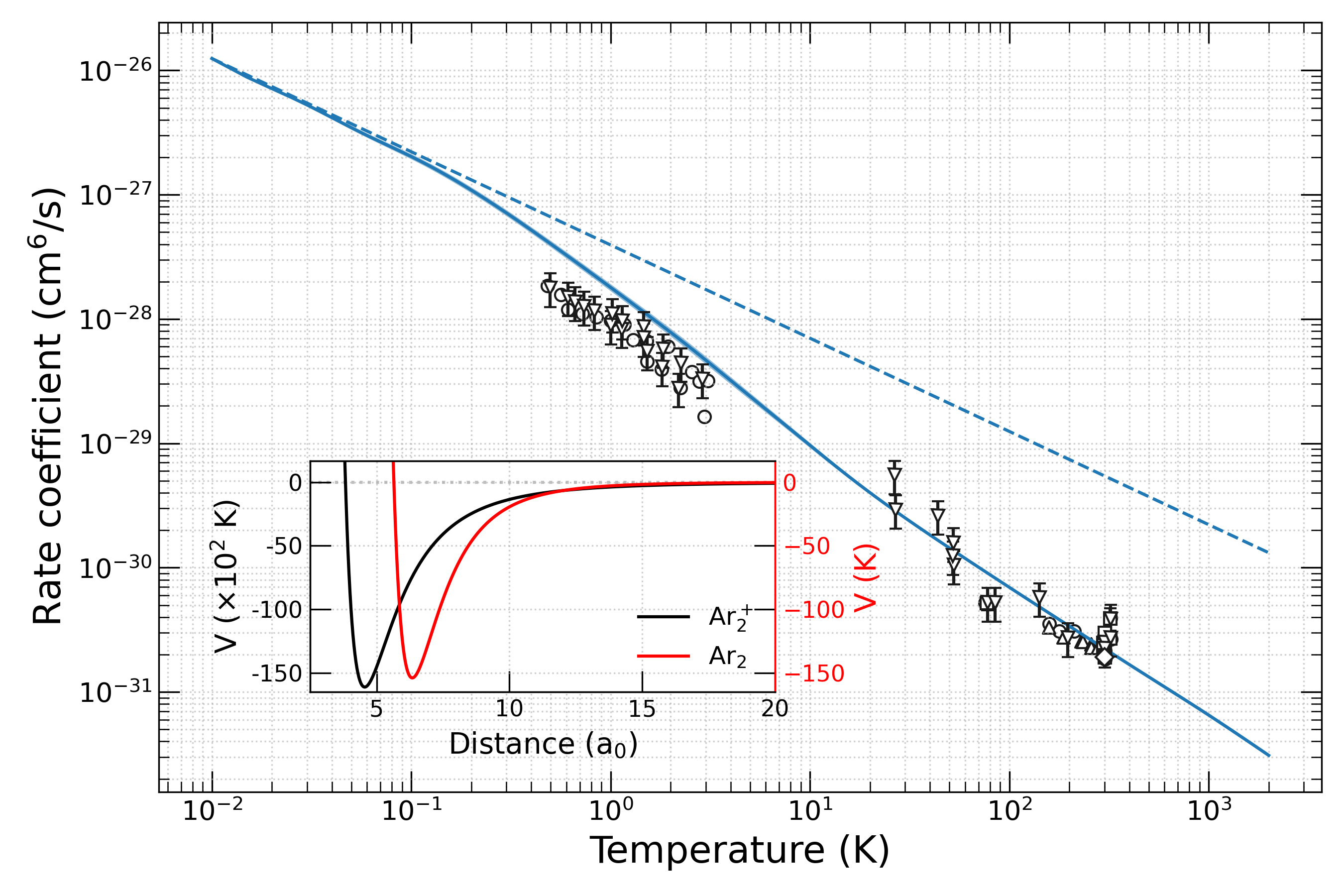}
    \caption{Temperature-dependent formation rate coefficient of \ce{Ar2+} via three-body recombination. The ion-atom recombination rate coefficient using a direct approach are presented in blue, with the shaded region representing the statistical error. The blue dashed line corresponds to the $T^{-3/4}$ classical threshold law shifted to match the prediction at low temperature. The experimentally determined rate coefficients are of 
    Ref.~\cite{Hawley_Smith_1992} (circles),
    Ref.~\cite{johnsenThreebodyAssociation1980a} (squares),
    Ref.~\cite{jonesTemperatureDependence1980} (triangles),
    Ref.~\cite{Hamon_Mitchell_Rowe_1998}(upside-down triangles) ,
    Ref.~\cite{Papanyan_Nersisyan_Ter-Avetisyan_Tittel_1995} (diamonds). The inset shows the Lennard-Jones potentials used to describe the \ce{Ar2+} (black axis) and \ce{Ar2} (dark red axis) interactions.}

    \label{fig:k3_ar}
\end{figure}

In Figure~\ref{fig:k3_ar}, the ion-atom recombination rate coefficient is presented from $T$ = 0.01~K to 2000~K, along with experimentally determined values. The predicted rates via the direct mechanism show excellent agreement with both the low-temperature and high-temperature experimental results. In addition, we see that at temperatures below 0.1~K the rate coefficient follows the $T^{-3/4}$ classical threshold law, which is dictated by the long-range charge-induced dipole interaction of \ce{Ar2+}. The pairwise interactions for \ce{Ar2+} and \ce{Ar2} dictate the temperature dependence of the reaction, and are shown in the inset.

\section{Conclusion}

We have shown that ion–atom recombination is governed by a direct three-body mechanism, without the need for intermediate complexes or steady-state assumptions. This result resolves longstanding discrepancies between theory and experiment and challenges the traditional Lindemann–Hinshelwood framework. More broadly, it establishes a unified description of barrierless termolecular reactions, opening new directions for understanding chemical kinetics across a wide range of environments.

Previous results on sulfur recombination~\cite{Sulfur}, halogen atom recombination~\cite{Koots2025}, ion-atom recombination in the cold regime~\cite{Krukow2016} and ozone recombination~\cite{Ozone}, support the same conclusion: barrierless termolecular reactions are mostly the result of the direct mechanism. Therefore, the direct mechanism should be incorporated as a main contributor in the study of the kinetics of termolecular reactions.

\bibliographystyle{rsc}
\bibliography{biblio}

\end{document}